\documentclass[useAMS,fleqn,usenatbib]{mnras}

\usepackage{graphicx}

\usepackage[T1]{fontenc}
\usepackage{ae,aecompl}
\usepackage{hyperref}
\newcommand{\hii}{H\,{\sc ii}\rm}

\newcommand{\nii}{[N\,{\sc ii}]}

\newcommand{\oiii}{[O\,{\sc iii}]}
\newcommand{\oii}{[O\,{\sc ii}]}

\newcommand{\sii}{[S\,{\sc ii}]}

\newcommand{\eg}{{e.g.}}

\newcommand{\lin}{$\,\lambda$}
\newcommand{\llin}{$\,\lambda\lambda$}

\newcommand{\oh}{12\,+\,log(O/H)}
\newcommand{\ohtwo}{\mbox{12\,+\,log(O/H)\,=\,}}

\newcommand{\halpha}{H$\alpha$}
\newcommand{\hbeta}{H$\beta$}
\newcommand{\hgamma}{H$\gamma$}
\newcommand{\nodata}{...}

\title[Abundance gradients in LSB galaxies]{Abundance gradients in low surface brightness  spirals: clues on the origin of common gradients in galactic discs
}
\author[F.~Bresolin and R.C. Kennicutt]{F.~Bresolin$^{1}$\thanks{E-mail:
bresolin@ifa.hawaii.edu} and R.C.~Kennicutt$^{2}$ 
\\
$^{1}$Institute for Astronomy, 2680 Woodlawn Drive, Honolulu, HI 96822, USA\\
$^{2}$Institute of Astronomy, University of Cambridge, Madingley Road, Cambridge CB3 0HA, UK\\
}

\begin{document}
\label{firstpage}
\pagerange{\pageref{firstpage}--\pageref{lastpage}}
\maketitle

%\label{firstpage}

\begin{abstract}
\noindent
We acquired spectra of 141 \hii\ regions in ten late-type low surface brightness galaxies (LSBGs). The analysis of the chemical abundances obtained from the nebular emission lines
shows that metallicity gradients are a common feature of LSBGs, contrary to previous claims concerning the absence of  such gradients in this class of galaxies. The average slope, when expressed in
units of the isophotal radius, is found to be significantly shallower in comparison to galaxies of high surface brightness. This result can be attributed to the reduced surface brightness range measured across
their discs, when combined with a universal surface mass density--metallicity relation. With a similar argument we explain the common abundance gradient observed in high surface brightness galaxy (HSBG) discs and its approximate dispersion. This conclusion is reinforced by our result that LSBGs share the same common abundance gradient  with HSBGs, when the slope is expressed in terms 
of the exponential disc scale length.

\end{abstract}


%==============================================================================================================
\section{Introduction}

During the past few decades the investigation of  the radial chemical abundance gradients of spiral galaxies has provided critical constraints for our understanding of their evolutionary histories. 
The overall picture that emerges from a large body of observational work  on the oxygen abundances (O/H) of extragalactic \hii\ regions, the standard tracers of metals in star-forming galaxies (see \citealt*{Pilyugin:2014} for a recent compilation), is consistent with an inside-out growth of galactic discs, in which the abundance gradients reflect  radial variations of  star formation rates and   gas infall timescales (\citealt{Prantzos:2000, Fu:2013}).

Recent observational work on the chemical composition of spiral galaxies has addressed a number of topics, in particular
the use of   abundance diagnostics that are complementary to \hii\ regions, such as blue supergiant stars (\citealt{Kudritzki:2008, Kudritzki:2012, Bresolin:2009a}) and planetary nebulae (\citealt{Bresolin:2010, Stasinska:2013}), azimuthal variations (\citealt*{Li:2013}; \citealt{Sanchez:2015}),
the flattening of the gradients at large galactocentric distances (\citealt{Bresolin:2009}; \citealt*{Bresolin:2012}; \citealt{Marino:2012})
and in interacting systems (\citealt{Kewley:2010, Rich:2012}), and the  spatial distributions of metals in high-redshift galaxies (\citealt{Yuan:2011, Queyrel:2012, Jones:2013}). Moreover, integral field spectroscopy is currently providing abundance gradients for  large  samples of nearby galaxies (\citealt{Rosales-Ortega:2010, Sanchez:2014}), yielding high statistical significance to correlations with global galactic properties and
scaling relations.

Despite this progress, the abundance gradient properties of low surface brightness galaxies (LSBGs), a significant component of the overall galaxy population (\eg\ \citealt{Impey:1997}), remain virtually unknown, largely as a result of observational challenges. Chemical abundances of late-type LSBGs have insofar been obtained either from emission line spectroscopy of the central portions of compact systems (\citealt*{Burkholder:2001}; \citealt*{Bergmann:2003}; \citealt{Liang:2010}) or from individual \hii\ regions, but with insufficient numbers 
to derive reliable radial trends (\citealt{McGaugh:1994, Roennback:1995}; \citealt*{Kuzio-de-Naray:2004}). To date the only attempt to constrain  
the oxygen abundance gradient in LSBGs has been reported by \citet{de-Blok:1998}, who, examining nebular abundances in three LSBGs, 
concluded  that their radial  distributions are flat.

The typical oxygen abundances of LSBGs span a wide range, from \oh\ $<$ 8.4 (\citealt{McGaugh:1994, Roennback:1995})
up to super-solar values\footnote{We adopt \oh$_\odot = 8.69$ from \citet{Asplund:2009}.}
 (\citealt{Bergmann:2003}), but on average peaking  below the characteristic abundances measured in
high surface brightness galaxies (\citealt{Liang:2010}). For example, \citet{Burkholder:2001} found a mean value \ohtwo 8.22 for 17 LSBGs, compared to \ohtwo 8.47 for 49 high surface brightness galaxies. 

The relatively low metallicities, together with  small star formation rates and blue optical and near-infrared colors,  corroborate the notion that LSBGs are slowly-evolving systems (\citealt*{de-Blok:1995}; \citealt{van-den-Hoek:2000}), with small-amplitude bursts  on top of star formation rates that are nearly constant when integrated over Gyr timescales (\citealt{Gerritsen:1999, Bell:2000, Boissier:2008, Schombert:2014}).
The reported lack of chemical abundance gradients has been regarded as supporting evidence for star formation histories characterized by sporadic episodes,  and possibly linked to a mode of galactic evolution that does not proceed from the inside out
(\citealt{de-Blok:1998, Vorobyov:2009}).

The present work stems from one main motivation. The study by \citet{de-Blok:1998},
which remains to date the only investigation of the radial abundance trends in LSBGs, reached its conclusion about the lack of detectable gradients
on the basis of only three galaxies, with a relatively small number ($4 \leq n \leq 7$) of \hii\ region spectra obtained per galaxy.
Perhaps more importantly, all three systems have an irregular morphological type. Abundance gradients in such galaxies are generally found to be very weak or absent (\eg\ \citealt{van-Zee:2006a, Croxall:2009}), although \citet*{Pilyugin:2015} recently measured significant gradients in a number of irregular galaxies.

Therefore, in order to verify claims, based on the chemical abundance properties, that LSBGs evolve differently (aside from the slower evolution) compared to galaxies of high surface brightness, it is 
essential to characterize the abundance gradients of LSBGs displaying a spiral morphology. 
Currently, multi-object techniques on 8m-class telescopes make the spectroscopy of relatively large samples of \hii\ regions, necessary to obtain reliable estimates of the abundance gradients, feasible even with moderate amounts of observing time. 

%Secondly,  recent discussions have suggested an analogy between the low surface brightness, extended discs of spiral galaxies and LSBGs, both in terms of their star formation properties (\citealt{Gil-de-Paz:2007, Thilker:2007, Boissier:2008}) and chemical abundance distributions (\citealt{Bresolin:2009}). This intriguing possibility calls for new observational constraints on the chemical abundance properties of  LSBGs.

In this paper we present new oxygen abundances of \hii\ regions in 10 late-type LSBGs, with the main goal of deriving their radial distributions. Our observational material is presented in Sect.~2, and we derive nebular chemical abundances using different diagnostics in Sect.~3. We  discuss the abundance gradients in Sect.~4 and   summarize our results in Sect.~5.

%==============================================================================================================
\section{Observations and data reduction}

\subsection{Observations}

\begin{table*}
 \centering
 \begin{minipage}{17.5cm}
  \centering
  \caption{Galaxy sample.}\label{sample}
  \begin{tabular}{lccccccccccc}
  \hline

ID	& R.A. (J2000)	         	& Dec. (J2000)                 	& Type 	& D   	& $M_B$ 	& PA  	& i 		& $r_{25}$		& $r_d$				& $\mu_B(0)$ 					& Ref.	\\
   	& h\,\, m\,\, s  	& $\circ$\,\,  $\prime$\,\, $\prime\prime$	&      		& Mpc 	&       	& \multicolumn{2}{c}{deg} 	& \multicolumn{2}{c}{arcsec (kpc)}  	& mag\,arcsec$^{-2}$	& \\
(1) & (2) & (3) & (4) & (5) & (6) & (7) & (8) & (9) & (10) & (11) & (12) \\
 \hline
ESO\,059-009 & 07 36 13.08	& $-$70 42 40.2  &	Sd		& 16.7	& $-$17.7 &	169	& 27 & 49 (3.9) & 27 (2.1) & 23.1 	& a\\
ESO\,187-051 & 21 07 32.78	& $-$54 57 09.1  &	Sm		& 17.8	& $-$16.8 &	11	& 52 & 40 (3.5) & 26 (2.2) & 23.1 	& 1,b \\
ESO\,249-036 & 03 59 15.72	& $-$45 52 20.7  &	Sm		& 9.7		& $-$14.9 &	156	& 15 & 35 (1.7) & 37 (1.7) & 24.0 	& 1,b \\
ESO\,440-049 & 12 05 34.07	& $-$31 25 25.4  &	Sc		& 31.2	& $-$19.0 &	62	& 42 & 45 (6.9) & 18 (2.7) & 23.1  	& 2,a\\
ESO\,446-053 & 14 21 17.08	& $-$29 15 47.6  &	Sdm		& 17.1	& $-$17.9 &	165	& 67 & 53 (4.4) & 28 (2.3)	& 22.9$^\star$  & 3,c \\
ESO\,504-025 & 11 53 50.15	& $-$27 20 59.2  &	Sd		& 24.1	& $-$17.5 &	20	& 40 & 21 (2.5) & 20 (2.3)	& 23.9$^\star$  & 3,c \\
ESO\,510-026 & 13 57 35.77	& $-$25 47 27.4  &	Scd		& 33.4	& $-$17.8 &	130	& 37 & 31 (5.1) & 13 (2.1)	& 22.4$^\star$  & 3,c \\
UGC\,1230    & 01 45 32.50	& $+$25 31 14.8  &	Sc		& 57.0	& $-$17.6 &	112	& 22 & 24 (6.5) & 16 (4.4)	& 23.3 	& 4,5,7,d \\
UGC\,6151    & 11 05 56.27	& $+$19 49 34.9  &	Sc-Sm	& 21.3	& $-$16.9 &	175	& 18 & 39 (4.1) & 24	(2.5) & 23.2 	& 6,5,7,d\\
UGC\,9024    & 14 06 40.55	& $+$22 04 12.4  &	Sc-Sm	& 36.5	& $-$16.4 &	148	& 15 & 18 (3.3) & 35 (6.2)	& 24.5 	& 5,7,d\\

 \hline
\end{tabular}
\end{minipage}
\begin{minipage}{17.5cm}
$(1)$ PA, i from \citet{Auld:2006}. 
$(2)$ $\mu_B(0)$ from \citet{Beijersbergen:1999}, using $\mu_R(0)$ and the average $\mu_B(0)-\mu_R(0)$ in their Table~2.
$(3)$ PA, i, Type from \citet{Matthews:1997}.
$(4)$ PA, i from \citet{van-der-Hulst:1993}.
$(5)$ Type from \citet{McGaugh:1995}.
$(6)$ PA, i from \citet{Patterson:1996}.
$(7)$ $r_d$ from \citet{McGaugh:1994a}.\\
Sources for the B-band exponential disc scalelength $r_d$: (a) \citet{Beijersbergen:1999} (R-band for ESO\,440-049); (b) \citet{Bell:2000}; (c) \citet{Lauberts:1989}; (d) \citet{McGaugh:1994a}.\\
$^\star$ Not a published value, but derived from the expression $\mu_o(B) = 25 - r_{25}/r_d$.
\end{minipage}
\end{table*}

LSBGs are often characterized by sparse distributions of faint \hii\ regions, which make it difficult 
to derive reliable chemical abundance gradients. For our study 
we selected  galaxies that were shown by previous work to contain a sufficient number of \hii\ regions to alleviate such a difficulty.
Our targets were extracted from a variety of sources providing imaging and surface photometry information: \citet*{Beijersbergen:1999}, \citet{Auld:2006}, \citet{de-Blok:1995}, \citet{McGaugh:1994a} and \citet*{McGaugh:1995}. The galaxies selected  have $B$-band central disc surface brightness $\mu_B(0) > 23$ mag\,arcsec$^{-2}$, a value that is commonly adopted to define LSBGs  (\eg\ \citealt{Impey:1997}), being more than one magnitude fainter than the canonical \citet{Freeman:1970} value  $\mu_B(0) =  21.65 \pm 0.3$ mag\,arcsec$^{-2}$ (this is somewhat arbitrary, and some authors use, for instance, $\mu_B(0) = 22$ mag\,arcsec$^{-2}$ as the dividing line between low and high surface brightness galaxies, HSBGs; \eg\ \citealt{Burkholder:2001,Boissier:2003,Liang:2010}).
We also included three targets from the list of extreme late-type spiral galaxies by \citet{Matthews:1997}. Although the value of their central disc surface brightness is not reported in their work, 
these systems have structural and global surface brightness characteristics that are comparable to those of the remaining targets in our sample. Their  $\mu_B(0)$ values  were estimated 
from the expression $\mu_B(0) = 25 - r_{25}/r_d$, using the  values for $r_{25}$ and $r_d$ from  \citet{Lauberts:1989}, and range between 22.4 and 23.9 mag\,arcsec$^{-2}$.

Our final sample of 10 LSBGs  comprises systems both with and without bulge, and
is presented in Table~\ref{sample}, where most of the information has been extracted from the HyperLeda database\footnote{http://leda.univ-lyon1.fr/} (\citealt{Makarov:2014}). The exceptions are reported in the last column of the table. The celestial coordinates of the galactic centers were measured from our  astrometrically calibrated images (see below). We follow the nomenclature from the ESO/Uppsala survey (\citealt{Lauberts:1982})
and the Uppsala General Catalogue of galaxies (UGC, \citealt{Nilson:1973}).
We re-classified  ESO\,249-036 (Horologium dwarf, part of a pair with ESO\,249-035) as an Sm type (instead of Im; \eg\ \citealt{Lauberts:1982}), based on the patchy, arm-like structure shown in the images by \citet[\citealt{Longmore:1982} classified this galaxy as an Sdm]{Auld:2006}.
For all the remaining galaxies we retain the late-type spiral (Sc to Sm) classification given in the sources specified in Table~\ref{sample}.

We obtained spectra of individual \hii\ regions in the target galaxies with two different instruments and multi-slit setups: the Very Large Telescope (VLT) equipped with the  FOcal Reducer and low dispersion Spectrograph (FORS)  and the Gemini North telescope with the Multi-Object
Spectrograph (GMOS). 
The  FORS data (collected for the seven ESO galaxies)
were obtained with the 300V grism %(central wavelength 5900~\AA) 
using 0\farcs 9 slits, while the  GMOS data (for the three UGC galaxies)
were obtained with the B600 grating %(central wavelengths 4800~\AA\ and 5700~\AA) 
using 
1\farcs 4 slits. %Our observations span a time range of about 20 months, as summarized in Table~\ref{log}. 

The slit masks were designed based on narrow-band H$\alpha$ imaging (on- and off-band) secured in advance of the spectroscopic runs.
For six of the galaxies we used two separate masks, in order to increase the number of target \hii\ regions. 
The number of 
\hii\ regions observed per galaxy ranges between 7 and 21. The total exposure times, reported in Table~\ref{log}, range between 
3600\,s and 6000\,s, and were subdivided into a number of sub-exposures varying from three to six. 
The spectral coverage extends from approximately 3650~\AA\ to 7000~\AA, with a FWHM spectral resolution of 8~\AA\ (FORS) and 4.8~\AA\ (GMOS). Thus all the main nebular emission-line diagnostics, from \oii\lin3727 to \sii\llin 6717,31, are accessible, and \nii\lin6583 is easily deblended from H$\alpha$.

\subsection{Data reduction}
Bias subtraction, flat-field correction and wavelength calibration were accomplished with the European Southern Observatory's EsoRex pipeline for the FORS data, and {\sc iraf}\footnote{{\sc iraf} is distributed by the National Optical Astronomy Observatories, which are operated by the Association of Universities for Research in Astronomy, Inc., under cooperative agreement with the National Science Foundation.}
routines in the {\tt\small gemini/gmos} package for the GMOS data. 
Flux calibration using standard star spectrophotometry, coaddition of sub-exposures and  extraction to one-dimensional spectra
 yielded the final data products to be analyzed.
Emission line fluxes were measured non-interactively with the {\tt\small fitprofs} tool in {\sc iraf}.

%NEW
The line fluxes were corrected for interstellar reddening by comparing the \halpha/\hbeta\ and \hgamma/\hbeta\  ratios to the corresponding case B values at $10^4$~K, requiring that the two line ratios yield a consistent extinction coefficient c(\hbeta), by adjusting the equivalent width of the underlying absorption component. 
The reddening-corrected line fluxes for the \oii\lin3727, \oiii\lin5007, \nii\lin6583 and \sii\llin6717,6731 emission lines, that are used as metallicity diagnostics,
 are reported, in units of $f$(\hbeta)\,=\,100, in Table~\ref{fluxes1} (the printed version reports the information for ESO\,059-009 only, while
 the complete table is available in the online version of the journal). We  include  the detections of the \oiii\lin4363 line, which can be used, in combination with \oiii\lin5007, to obtain direct estimates of the gas electron temperature. The errors quoted include the statistical and flux calibration errors,  and the uncertainty in the reddening correction.
We confirmed that the relative strengths of the emission lines are consistent with those expected for photoionized nebulae, as judged from the appropriate diagnostic diagrams (\eg\ \citealt*{Baldwin:1981}).

%  TABLE: OBSERVATIONS LOG
\begin{table}
 \centering
  \caption{Log of the observations.}\label{log}
  \begin{tabular}{lcc}
  \hline

%VLT - FORS & &  \\
%\multicolumn{3}{l}{Grism 300V, $\lambda_c = 5900$~\AA, 0\farcs 9 slits} \\[1mm]
Galaxy       & Date           & Exp. time (s) \\
\hline
ESO\,059-009 & 2012 Apr 17,18     & 6000 ($\times$2 fields) \\
ESO\,187-051 & 2011 June 25,26	 & 5094  \\
ESO\,249-036 & 2011 Mar 04,05	 & 5094  \\
ESO\,440-049 & 2012 Apr 17     & 6000 ($\times$2 fields)\\
ESO\,446-053 & 2012 Apr 17,18  & 3600 ($\times$2 fields)   \\
ESO\,504-025 & 2012 Apr 18     & 6000 ($\times$2 fields)  \\ 
ESO\,510-026 & 2012 Apr 17     & 6000  \\[1mm]
%\hline

%Gemini N - GMOS & &  \\
%\multicolumn{3}{l}{Grating B600-G5307, $\lambda_c = 4800, 5700$~\AA, 1\farcs 4 slits} \\[1mm]
%Galaxy       & Date           & Exp. time (s) \\
%\hline
UGC\,1230 & 2010 Sept 11,12       & 3600 ($\times$2 fields) \\
UGC\,6151 & 2011 May 04       & 4500  \\
          & 2012 Mar 20       & 4500  \\
UGC\,9024 & 2011 June 20,30       & 4500  \\
          & 2012 Feb 26, Apr 15    & 4500  \\
\hline
\end{tabular}
\end{table}
% ..................................................................................................

% ..................................................................................................................
%  TABLE: FLUXES 1
\begin{table*}
 \centering
  \begin{minipage}{16.cm}
  \centering
  \caption{Reddening-corrected fluxes: ESO\,059-009}\label{fluxes1}
  \begin{tabular}{lcccccccccc}
  \hline

ID	&	R.A.		&	Dec.		&	$r/r_{25}$		&	\oii\		&	\oiii\ 		&	\oiii\		&	\nii\		&	\sii\			&	c(H$\beta$)	&	F(H$\beta$)	\\
	&	(J2000.0)	&	(J2000.0)	&				&	3727  	&	4363		&	5007		&	6583		&	6731+6717	&				&	(erg s$^{-1}$ cm$^{-2}$)			\\
(1) & (2) & (3) & (4) & (5) & (6) & (7) & (8) & (9) & (10) & (11)  \\
 \hline
 1 & 07 36 12.4  & $-70$ 42 03.50  & 0.76 &    179 $\pm$    9 &     \nodata        &      33 $\pm$    2 &     71 $\pm$    4 &      76 $\pm$    3 &   0.21 &     1.5 $\times\ 10^{-16}$ \\  %slit103
 2 & 07 36 08.1  & $-70$ 42 09.10  & 0.84 &    190 $\pm$   10 &     \nodata        &      37 $\pm$    2 &     67 $\pm$    4 &      70 $\pm$    3 &   0.26 &     1.1 $\times\ 10^{-15}$ \\  %slit104
 3 & 07 36 10.6  & $-70$ 42 10.62  & 0.66 &    172 $\pm$    9 &     \nodata        &      45 $\pm$    2 &     69 $\pm$    4 &      51 $\pm$    2 &   0.46 &     1.5 $\times\ 10^{-16}$ \\  %slit203
 4 & 07 36 13.1  & $-70$ 42 24.78  & 0.32 &    116 $\pm$    7 &     \nodata        &       0 $\pm$    1 &     70 $\pm$    4 &      66 $\pm$    3 &   0.43 &     7.9 $\times\ 10^{-17}$ \\  %slit204
 5 & 07 36 05.9  & $-70$ 42 26.15  & 0.86 &    164 $\pm$    8 &     \nodata        &      54 $\pm$    2 &     66 $\pm$    4 &      59 $\pm$    2 &   0.30 &     1.3 $\times\ 10^{-15}$ \\  %slit106
 6 & 07 36 15.5  & $-70$ 42 26.62  & 0.40 &    168 $\pm$   10 &     \nodata        &      12 $\pm$    1 &     72 $\pm$    4 &      52 $\pm$    2 &   0.51 &     9.4 $\times\ 10^{-17}$ \\  %slit105
 7 & 07 36 15.0  & $-70$ 42 38.19  & 0.23 &    119 $\pm$    6 &     \nodata        &      18 $\pm$    1 &     82 $\pm$    4 &      44 $\pm$    2 &   0.26 &     1.6 $\times\ 10^{-16}$ \\  %slit205
 8 & 07 36 05.3  & $-70$ 42 43.60  & 0.90 &    194 $\pm$   10 &     \nodata        &      51 $\pm$    2 &     61 $\pm$    3 &      53 $\pm$    2 &   0.45 &     2.4 $\times\ 10^{-16}$ \\  %slit108
 9 & 07 36 10.5  & $-70$ 42 44.25  & 0.31 &    166 $\pm$    9 &     \nodata        &      29 $\pm$    2 &     96 $\pm$    5 &      83 $\pm$    3 &   0.55 &     7.4 $\times\ 10^{-17}$ \\  %slit206
10 & 07 36 14.4  & $-70$ 42 46.88  & 0.20 &    111 $\pm$    6 &     \nodata        &      15 $\pm$    1 &     74 $\pm$    4 &      59 $\pm$    2 &   0.28 &     6.4 $\times\ 10^{-17}$ \\  %slit107
11 & 07 36 10.8  & $-70$ 42 54.22  & 0.40 &    188 $\pm$   10 &     \nodata        &      89 $\pm$    4 &     81 $\pm$    4 &      63 $\pm$    3 &   0.35 &     5.8 $\times\ 10^{-17}$ \\  %slit208
12 & 07 36 13.1  & $-70$ 42 57.41  & 0.35 &    187 $\pm$   10 &     \nodata        &      63 $\pm$    3 &     71 $\pm$    4 &      40 $\pm$    2 &   0.41 &     8.8 $\times\ 10^{-17}$ \\  %slit109
13 & 07 36 16.3  & $-70$ 43 07.06  & 0.65 &    165 $\pm$    8 &     \nodata        &      14 $\pm$    1 &     79 $\pm$    4 &      85 $\pm$    3 &   0.31 &     8.7 $\times\ 10^{-17}$ \\  %slit110
14 & 07 36 11.9  & $-70$ 43 09.27  & 0.62 &     96 $\pm$    8 &     \nodata        &       0 $\pm$    1 &     79 $\pm$    5 &      75 $\pm$    4 &   0.12 &     5.0 $\times\ 10^{-17}$ \\  %slit209
15 & 07 36 12.5  & $-70$ 43 13.57  & 0.70 &    193 $\pm$   10 &     \nodata        &      26 $\pm$    1 &     84 $\pm$    5 &      84 $\pm$    3 &   0.45 &     1.3 $\times\ 10^{-16}$ \\  %slit111
16 & 07 36 16.7  & $-70$ 43 36.63  & 1.22 &    286 $\pm$   18 &     \nodata        &      70 $\pm$    5 &     76 $\pm$    5 &      57 $\pm$    3 &   0.50 &     2.6 $\times\ 10^{-17}$ \\  %slit113
 \hline
\end{tabular}
\end{minipage}
\begin{minipage}{16cm}
The complete table, which includes information for the full sample, is available in plain text format in the online version.\\
F(H$\beta$) in column (11) is the measured flux of the H$\beta$ emission line.
\end{minipage}

\end{table*}
% ..................................................................................................................

%==============================================================================================================
\section{Chemical abundances and radial gradients}

Measuring nebular chemical abundances that are free of large systematic uncertainties remains an unsolved problem in astrophysics. The nature of this issue and the various 
attempts to circumvent the difficulties have been extensively covered by a large number of authors and will not be repeated here (see, for example, \citealt*{Bresolin:2004}; \citealt{Kewley:2008,Lopez-Sanchez:2012}). In essence, several emission line diagnostics and different calibrations 
of these diagnostics have been proposed in the literature, with systematic variations on the derived oxygen abundances that reach up to 0.7 dex. One additional complication that is especially relevant for the current study is the non-monotonic nature of some of the diagnostics, which can introduce large uncertainties when attempting to measure radial abundance gradients.

A frequently used strategy, and the one we adopt here, is to consider a number of different abundance determination methods, with the intention of identifying the quantities that are largely independent of the  diagnostics used. For example, radial abundance trends are generally found to be qualitatively invariant relative to the choice of abundance diagnostics, but one needs to be aware of the fact that different methods can yield different gradient slopes (\eg\ \citealt{Bresolin:2009a}). 
In order to compare our results with  recent studies concerning the statistical properties of the gradients of large samples of spiral galaxies, discussed below, we have considered the following nebular metallicity diagnostic methods (as is customary for nebular studies, we use the terms metallicity and oxygen abundance interchangeably):

\begin{enumerate}

\item\noindent O3N2 $\equiv$ log[(\oiii\lin5007/\hbeta)/(\nii\lin6583/\halpha)], with the empirical calibration given by \citet{Pettini:2004}, which allows a comparison  with the study of a large sample of spiral galaxies by \citet{Sanchez:2014}.\\

\item\noindent N2O2 $\equiv$ \nii\lin 6583/\oii\lin3727, adopting both the calibration based on photoionization models by \citet[=KD02]{Kewley:2002} and the empirical one by \citet[=B07]{Bresolin:2007}. The recent study of abundance gradients in local star-forming galaxies by \citet{Ho:2015} is based on this diagnostic. We have already shown in \citet{Bresolin:2009} that the  calibrations by KD02 and B07 yield abundance gradients having virtually the same slopes, despite a large systematic offset.\\

\item\noindent N2 $\equiv$ \nii\lin6583/\halpha, with the calibration by  \citet{Pettini:2004}.\\

\item\noindent $R_{23}$ $\equiv$ (\oii\lin3727 + \oiii\llin4959,5007)/\hbeta, adopting the calibration by \citet{McGaugh:1991}, in the analytical form given by \citet*{Kobulnicky:1999}. We used this diagnostic in order to check whether the metallicity gradients derived from the previous diagnostics, all involving the nitrogen \nii\lin6583 line, are confirmed by considering only oxygen  lines instead. Unfortunately, the use of this indicator for abundance gradient studies is complicated by the well-known non-monotonic behavior of $R_{23}$.
In order to attempt and break the degeneracy we followed other authors in using the \nii\lin6583/\oii\lin3727 line ratio: objects for which \nii\lin6583/\oii\lin3727 $>$ $-1.2$ should belong to the $R_{23}$ upper branch (\eg\ \citealt{Kewley:2008}). For our sample we found that, in order to obtain monotonic radial trends in abundance for most galaxies we need to arbitrarily vary  this boundary between $-1.3$ and $-1.0$. Even so, some of the radial trends obtained from $R_{23}$ display quite a large scatter or even sudden discontinuities, which are unphysical and largely due to the uncertainty in placing objects on the correct branch, and the fact that many of the \hii\ regions lie in the `turnaround' region of the diagnostic. For our purposes this is of secondary importance: we simply wish to demonstrate that the radial abundance trends we observe do not depend on the selection of diagnostics based on nitrogen lines instead of oxygen lines.
\end{enumerate}

\begin{figure*}
\center
\includegraphics[width=0.9\textwidth]{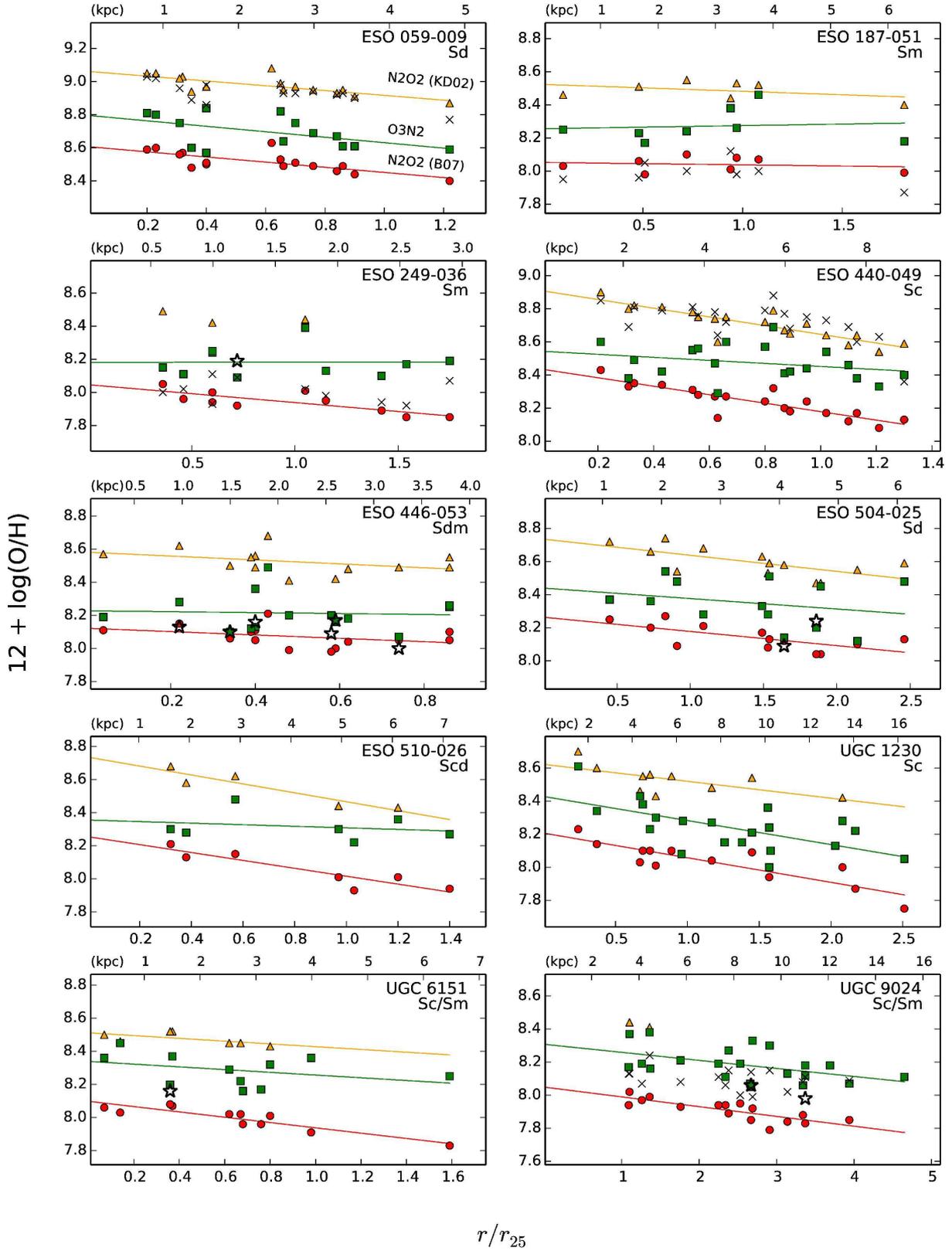}
\caption{Radial O/H abundance gradients drawn in terms of the isophotal radius $r_{25}$ (lower scale of each panel) and of the physical radius in kpc (upper scale) from different diagnostics (see symbol explanation in the top left panel). In addition, star symbols are used for abundances obtained with the direct method (when available), and crosses for abundances from $R_{23}$ (only shown when branch selection is unambiguous). The lines represent least square fits to the abundance data. 
See text for explanations.
\label{gradients}}
\end{figure*}

The radial O/H abundance  gradients we derive for our sample  are shown in Fig.~\ref{gradients}. The linear least square fits to the log(O/H) data obtained from the O3N2 and the N2 methods yield the same slopes, within the uncertainties, and we therefore plot only the results obtained using the former indicator, using green squares for the individual \hii\ regions, and the green line representing the least square fit to these data points. For the N2O2 diagnostic we show the data and the corresponding linear fits using both the B07 (red circles and line) and the KD02 (yellow triangles and line) calibrations, to illustrate the fact that they yield the same gradient slopes.
We used the KD02 polynomial expression for the N2O2 index vs. O/H relation calculated for a ionization parameter $q=2\times10^7$\,cm\,s$^{-1}$. The N2O2 diagnostic becomes insensitive to oxygen abundance for low O/H values. For the KD02 diagnostic we have included in Fig.~\ref{gradients}  only data points for which log(\nii\lin6583/\oii\lin3727)$ > -1.2$, corresponding 
approximately to \oh\ $>$ 8.4 in the KD02 abundance scale. Below such limit the KD02-based O/H values display large and sudden deviations from the smooth radial trends seen in Fig.~\ref{gradients}, symptomatic of the breakdown of the calibration at low metallicity.
We instead extended the use of the empirical B07 calibration to a lower limit, log(\nii\lin6583/\oii\lin3727)$ > -1.4$, 
corresponding approximately to \oh\ $>$ 7.8 in the B07 abundance scale. Therefore, the number of data points based on the B07 calibration exceeds that for the KD02 calibration for some galaxies (especially ESO\,249-036 and UGC~9024, for which we are not able to derive a gradient based on the KD02 calibration; we required a minimum of five data points for a fit to the abundance gradients), but we stress again that  the two calibrations yield slopes that are in very good agreement with each other.

In Fig.~\ref{gradients} we also include, using star symbols, the O/H values obtained from the direct method, i.e. based on the measurement of the electron temperature using the \oiii\lin4363/\lin5007 line ratio (this procedure was carried out with the {\tt nebular} package in {\sc iraf}; see \citealt{Bresolin:2009a} for details).
The weak \oiii\lin4363 auroral line was measured  for 12 \hii\ regions (six in ESO\,446-053). It can be seen that the direct abundances are in rough agreement with the O3N2 and N2O2 (B07) diagnostics, as expected, since both these calibrations are based on samples of \hii\ regions with direct abundance determinations. 
Finally, we include in Fig.~\ref{gradients} the O/H abundances obtained from  $R_{23}$ (cross symbols, without showing the corresponding linear fits), but only for those galaxies for which the selection between the upper and lower branches was  unequivocal, based on the \nii\lin6583/\oii\lin3727 ratio. In the case of the five galaxies for which, over the full radial range covered by the \hii\ regions,
this line ratio straddles the approximate boundary between the two branches [log(\nii\lin6583/\oii\lin3727) $\simeq$ $-1.2$; \eg\ \citealt{Kewley:2008})] we do not show the $R_{23}$-based abundances.

% ..................................................................................................................
% ..................................................................................................................

Table~\ref{tab-gradients}  and ~\ref{tab-zp} summarize the slopes and zero points, respectively, of the metallicity gradients obtained when the nebular O/H values are based on the O3N2 and the N2O2 (B07) diagnostics. The slopes are reported, together with their errors, in terms of the physical size of the galaxies in kpc, of the isophotal radius $r_{25}$ and of the $B$-band exponential disc scale length $r_d$. We enclose these values in brackets when we wish to identify galaxies for which 
the errors exceed the absolute values of the slope, i.e. when the abundance gradient is formally consistent with a flat radial distribution.
At the bottom of Table~\ref{tab-gradients}  we include the average slopes, with corresponding errors,  and their standard deviations, calculated including all ten galaxies.

\medskip
Before proceeding with the discussion, we make a few preliminary remarks, based on the results presented in Fig.~\ref{gradients} and Table~\ref{tab-gradients}:

\begin{enumerate}

\item\noindent  a large fraction (50 per cent using the O3N2 diagnostic, 90 per cent using N2O2) of the sample galaxies present a significant radial abundance gradient
%Between half and the vast majority of the galaxies in the sample, depending on the diagnostic used, have a significant radial abundance gradient. 
The higher number of galaxies with gradients that are compatible with being flat in the case of O3N2 (5 out of 10) is related to the next point. Only one galaxy, ESO\,187-051, has a flat gradient according to the N2O2 diagnostic.
\medskip

\item\noindent In Fig.~\ref{gradients} the scatter of the data points corresponding to O3N2 (the mean rms of the linear fit is 0.095 dex) is approximately two times larger than for N2O2 (mean rms=0.044). 
\medskip

\item\noindent The slopes derived using O3N2 are either comparable to (within the errors) or shallower (ESO\,249-036, ESO\,440-049, ESO\,510-026) than those obtained from N2O2. We interpret this difference and   (ii) above with the higher dependence of O3N2 on the ionization parameter (\eg\ \citealt{Ho:2015}), and the fact that this diagnostic is constructed with both low- and high-excitation emission lines.
While it is not immediately obvious why O3N2 should yield shallower gradients in some galaxies, we also point out that in absolute terms the  difference in slope that can be obtained between the two diagnostics, while statistically significant, is small in our sample of galaxies in relation to the range of values found in much larger sample of  spirals.
Considering, for example, gradients in units of disc scale lengths 
the average slope from  O3N2 is $-0.045\pm 0.013$ dex\,$r_d^{-1}$, compared to $-0.085\pm 0.010$ dex\,$r_d^{-1}$ from N2O2 (i.e.~the difference is $\Delta_{O3N2-N2O2} = 0.04$ dex\,$r_d^{-1}$).
The slopes found by \citet{Sanchez:2014} in a sample of 193 spiral galaxies (using O3N2) cover a much larger range, from  $-0.19$  to $+0.04$ dex\,$r_d^{-1}$.
Considering isophotal radii instead, the average slope from  O3N2 is $-0.065\pm 0.018$ dex\,$r_{25}^{-1}$, compared to $-0.133\pm 0.024$ dex\,$r_{25}^{-1}$  from N2O2
($\Delta_{O3N2-N2O2} = 0.07$ dex\,$r_{25}^{-1}$). For the bulk of the 49 galaxies studied by 
\citet{Ho:2015}  using N2O2 the slope ranges from 0 to $-0.8$ dex\,$r_{25}^{-1}$.  Therefore, the  small systematic difference resulting from the selection of the abundance diagnostic 
does not hamper the comparison with the distribution of abundance gradient properties of HSBGs carried out in Sect.4.

\medskip

\item\noindent In those cases where it has been possible to derive a reliable radial abundance trend also from $R_{23}$, we find results that are compatible either with N2O2 (all galaxies except ESO\,249-036)
 or O3N2 (all galaxies except ESO\,440-049).

\end{enumerate}

For the reasons outlined in points (ii) and (iii) above, it is tempting to assign more significance to the results obtained from N2O2. For completeness, in the following discussion we  will continue to consider gradients obtained from both  O3N2 and N2O2.

% ..................................................................................................................
%  TABLE: GRADIENTS
\begin{table*}
 \centering
  \caption{Abundance gradient slopes.}\label{tab-gradients}
  \begin{tabular}{l@{\hskip 0.2cm}c@{\hskip 0.2cm}cc@{\hskip 0.2cm}cc@{\hskip 0.2cm}c@{\hskip 0.8cm}c@{\hskip 0.2cm}cc@{\hskip 0.2cm}cc@{\hskip 0.2cm}c}
  \hline

ID	&	\multicolumn{6}{c}{O3N2}								&			\multicolumn{6}{c}{N2O2}			\\
	&	dex kpc$^{-1}$	&	$\sigma$  		&	dex $r_{25}^{-1}$	&	$\sigma$	&	dex $r_d^{-1}$		& 	$\sigma$	&	dex kpc$^{-1}$	&	$\sigma$  		&	dex $r_{25}^{-1}$	&	$\sigma$	&	dex $r_d^{-1}$		& 	$\sigma$	\\
 \hline
ESO 059-009   &  $-0.042$ & 0.020 & $  -0.165$ & 0.080 & $  -0.090$ & 0.044 &   $-0.040$ & 0.010 & $  -0.156$ & 0.039 & $  -0.085$ & 0.021 \\ 
ESO 187-051   &  ($+0.005$ & 0.023 & $  +0.019$ & 0.079 & $  +0.012$ & 0.051) &   ($-0.004$ & 0.010 & $  -0.014$ & 0.035 & $  -0.009$ & 0.022) \\ 
ESO 249-036   &  ($+0.001$ & 0.040 & $  +0.001$ & 0.066 & $  +0.001$ & 0.070) &   $-0.065$ & 0.018 & $  -0.108$ & 0.030 & $  -0.114$ & 0.031 \\ 
ESO 440-049   &  $-0.013$ & 0.011 & $  -0.092$ & 0.076 & $  -0.036$ & 0.030 &   $-0.037$ & 0.005 & $  -0.256$ & 0.035 & $  -0.101$ & 0.014 \\ 
ESO 446-053   &  ($-0.006$ & 0.031 & $  -0.028$ & 0.138 & $  -0.015$ & 0.073) &   $-0.023$ & 0.017 & $  -0.102$ & 0.074 & $  -0.054$ & 0.039 \\ 
ESO 504-025   &  ($-0.025$ & 0.028 & $  -0.063$ & 0.069 & $  -0.060$ & 0.065) &   $-0.034$ & 0.011 & $  -0.086$ & 0.028 & $  -0.081$ & 0.026 \\ 
ESO 510-026   &  ($-0.009$ & 0.017 & $  -0.047$ & 0.087 & $  -0.019$ & 0.035) &   $-0.047$ & 0.009 & $  -0.239$ & 0.047 & $  -0.097$ & 0.019 \\ 
UGC 1230      &  $-0.022$ & 0.007 & $  -0.145$ & 0.043 & $  -0.099$ & 0.029 &   $-0.023$ & 0.004 & $  -0.149$ & 0.026 & $  -0.101$ & 0.018 \\ 
UGC 6151      &  $-0.020$ & 0.017 & $  -0.082$ & 0.069 & $  -0.050$ & 0.042 &   $-0.039$ & 0.006 & $  -0.161$ & 0.026 & $  -0.098$ & 0.016 \\ 
UGC 9024      &  $-0.015$ & 0.006 & $  -0.049$ & 0.020 & $  -0.092$ & 0.038 &   $-0.018$ & 0.004 & $  -0.059$ & 0.012 & $  -0.112$ & 0.022 \\ 
\\[-2mm] \cline{2-13} \\ 
Average &$  -0.015$ & 0.004 &$  -0.065$ & 0.018 &$  -0.045$ & 0.013 &$  -0.033$ & 0.005 &$  -0.133$ & 0.024 &$  -0.085$ & 0.010 \\ 
Standard dev.        &\phantom{-}   0.013 &       &\phantom{-}   0.055 &       &\phantom{-}   0.038 &       &\phantom{-}   0.016 &       &\phantom{-}   0.072 &       &\phantom{-}   0.030 &       \\ 

 \hline
\end{tabular}
\begin{minipage}{17cm}
The brackets are used to identify regressions for which the slope is flat within the quoted errors.
\end{minipage}
\end{table*}
% ..................................................................................................................

% ..................................................................................................................
%  TABLE: ZERO POINTS
\begin{table*}
 \centering
  \caption{Abundance gradient zero points.}\label{tab-zp}
  \begin{tabular}{l@{\hskip 0.5cm}cc@{\hskip 1.2cm}cc}
  \hline

ID	&	\multicolumn{2}{c}{O3N2\phantom{aaaaaa}}	& 	\multicolumn{2}{c}{N2O2}			\\
        &      12+log(O/H)   & $\sigma$  & 12+log(O/H)   & $\sigma$ \\ 
 \hline
ESO 059-009   &     8.796 &  0.054 &    8.607 &  0.025 \\ 
ESO 187-051   &     8.256 &  0.076 &    8.052 &  0.033 \\ 
ESO 249-036   &     8.181 &  0.071 &    8.046 &  0.032 \\ 
ESO 440-049   &     8.543 &  0.062 &    8.434 &  0.029 \\ 
ESO 446-053   &     8.227 &  0.075 &    8.122 &  0.040 \\ 
ESO 504-025   &     8.439 &  0.105 &    8.263 &  0.043 \\ 
ESO 510-026   &     8.355 &  0.080 &    8.255 &  0.043 \\ 
UGC 1230      &     8.428 &  0.061 &    8.206 &  0.035 \\ 
UGC 6151      &     8.339 &  0.052 &    8.098 &  0.019 \\ 
UGC 9024      &     8.308 &  0.055 &    8.049 &  0.029 \\ 

 \hline
\end{tabular}
\end{table*}
% ..................................................................................................................

%==============================================================================================================
\section{Discussion}
The first  finding of this work has already  been introduced in the previous section: late-type (spiral) LSBGs {\em do} have measurable radial metallicity gradients. We thus do not confirm the earlier result
by \citet{de-Blok:1998}, who found no evidence for the presence of such gradients in LSBGs. The origin of this discrepancy can be attributed to a number of factors that are likely to affect their study:
small sample size (three galaxies), galaxy morphology (irregulars), small number of \hii\ regions with limited radial coverage, and possibly the use of the $R_{23}$ diagnostic in the uncertain turnaround region.
We believe that our result is robust, being based on a relatively large number of \hii\ regions per galaxy (up to 21), and
on  deeper spectra than was possible in the 1990's.

%This contradicts the qualitative statement made in the pioneering work by \citet{de-Blok:1998} on the basis of \hii\ region data for three LSBGs,  that  established the  long-standing paradigm concerning the metallicity gradients of LSBGs present in the literature  until now. 

Our study finally answers the question raised by \citet{Edmunds:1999}, who speculated that the chemical evolution of low and high surface brightness galaxies should proceed in a similar fashion,
and wondered whether LSBGs and HSBGs `might {\em not} show similar radial abundance gradients, since organization of star formation by spiral structure is a good candidate mechanism for gradient generation'. Indeed our observations show that, as in the case of HSBG spirals, metallicity gradients are a common characteristic of spiral LSBGs,  implying that models of  inside-out disc formation  can explain the chemical
abundance properties of these galaxies, too. It is not necessary to invoke drastically different modes of disc buildup for LSBGs, as suggested in the literature in order to account for the apparent lack of abundance gradients (\citealt{de-Blok:1998, Vorobyov:2009}).
\medskip

% We also note that, focussing on the chemical abundance analysis based on N2O2, the only galaxy with a flat gradient, ESO\,187-051, has a late, Sm morphological type. Earlier galaxy types (Sc and Sd), but also other Sm's,  all appear to have gradients in metallicity. This might suggest that the presence (and  strength?) of spiral structure, rather than surface brightness, is the principal factor contributing to the the metallicity gradients we observe in star-forming galaxies.

The  question we wish to address next is whether  the gradients observed in LSBGs are quantitatively comparable to those of HSBGs. For this purpose we compare the slopes obtained in Section~3 with those measured from  large samples
of HSBGs recently published  by \citet{Sanchez:2014} and \citet{Ho:2015}. Both studies confirm, with higher statistical significance,  earlier findings obtained from smaller samples of galaxies and \hii\ regions (\eg\ \citealt{Vila-Costas:1992}; \citealt*{Zaritsky:1994}; \citealt{Garnett:1997}), namely
 that spirals are characterized by similar abundance gradients, irrespective of morphological type and other global properties, when normalized to either the exponential disc scale length or the isophotal radius.
\medskip

\begin{figure}
\center
\includegraphics[width=0.47\textwidth]{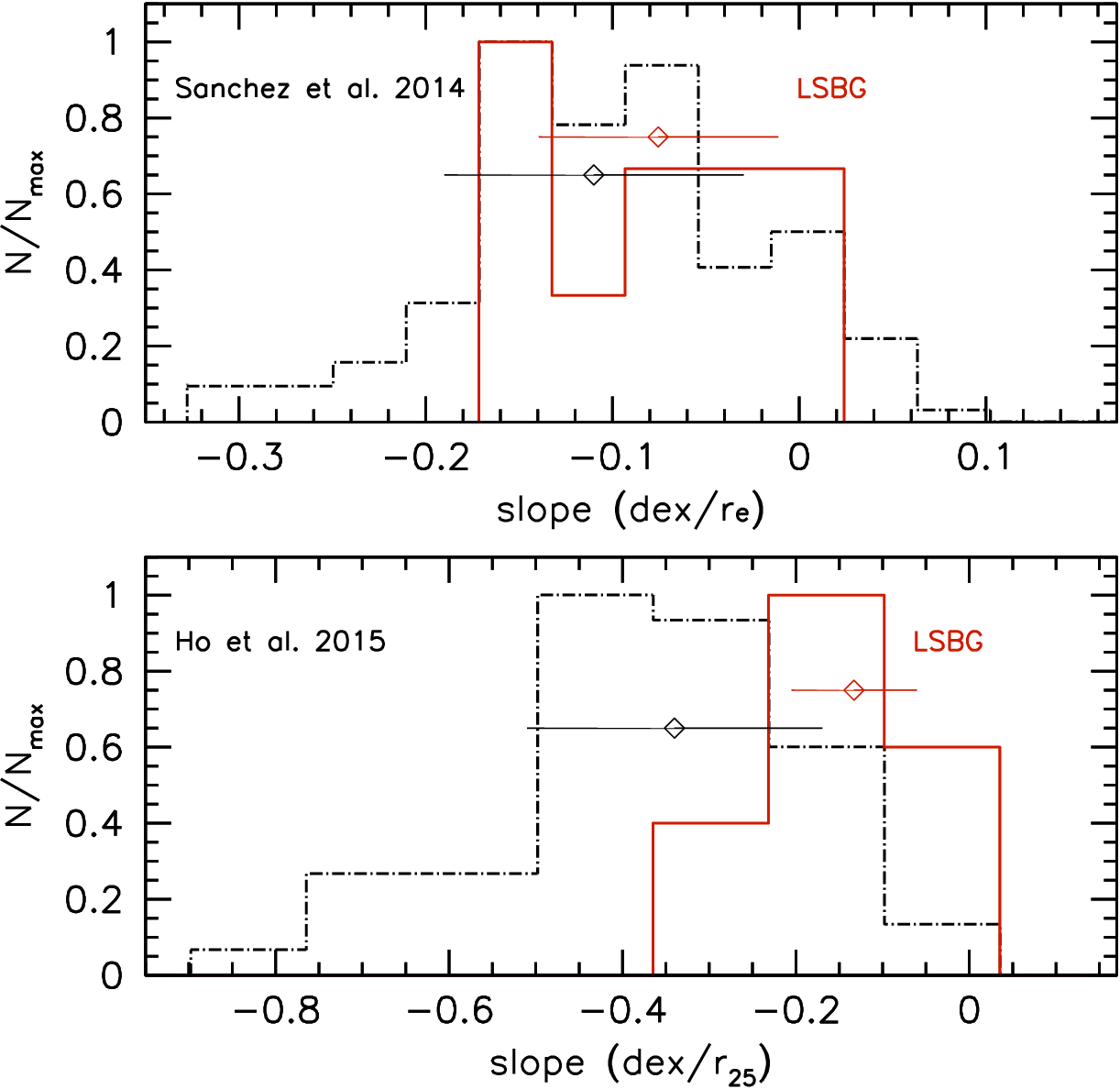}
\caption{{\em (Top)} Histogram of the slope distribution for the LSBGs (red continuous line) and for the sample by \citet[black dot-dashed line]{Sanchez:2014}. Units for the slope are dex\,$r_e^{-1}$.
{\em (Bottom)} Histogram of the slope distribution for the LSBGs (red continuous line) and for the sample by \citet[black dot-dashed line]{Ho:2015}. Units for the slope are dex\,$r_{25}^{-1}$.
In both panels mean values (diamonds) and standard deviations (horizontal bars) are indicated for the different samples.
\label{histogram}}
\end{figure}

\noindent
{\em a})
\citet{Sanchez:2014} analyzed the abundance gradient properties of 306 galaxies observed as part of the Calar Alto Legacy Integral Field Area survey (CALIFA, \citealt{Sanchez:2012a}). 
Chemical abundances could be derived for at least four \hii\ regions per system in 193 galaxies, allowing  a determination of their radial abundance gradients.
These authors established the existence of a characteristic slope of the  radial metallicity (O/H) gradient, that is independent of the properties of non-interacting, isolated galaxies, such as morphological type, presence of bars, and stellar mass. Considering only their subsample of 146 isolated galaxies, showing no evidence for ongoing interactions, the characteristic gradient,
 expressed in terms of the disc effective radius $r_e$, has a value $\alpha = -0.11$ dex\,$r_e^{-1}$, with a standard deviation of 0.08 dex\,$r_e^{-1}$. The normalized histogram of the slope distribution, extracted from  Fig.~8 in \citet{Sanchez:2014}, is shown in the top panel of Fig.~\ref{histogram} by the black, dot-dashed line.
 
We can make a direct comparison with the average value we obtained for LSBGs using the O3N2 diagnostic, which was also adopted by \citet{Sanchez:2014}: $\alpha_{r_d} = -0.045 \pm 0.038$ dex\,$r_d^{-1}$ (mean $\pm$ standard deviation), which translates into $\alpha_{r_e} = -0.076 \pm 0.064$ dex\,$r_e^{-1}$ (accounting for the fact that $r_e = 1.678\,r_d$).
The distribution of the slope values we obtain for the LSBGs is shown in the top panel of Fig.~\ref{histogram} by the red, continuous line. It can be seen that
the mean slopes for the HSBGs and LSBGs are indistinguishable, and that the two  distributions are also compatible with each other, as confirmed by the $t$--test and the Kolgomorov--Smirnov test, respectively.

We note that, while the CALIFA sample analyzed by \citet{Sanchez:2014}  is complete only at the bright end, the luminosity of the 146 non-interacting galaxies that constitute our comparison sample
extends down to $M_g \simeq -17$, thus overlapping with the luminosity range of the LSBGs in our sample. Moreover, the analysis by \citet{Sanchez:2014} excluded trends of abundance gradient slopes with galaxy luminosity.

% ..................................................................................................................
% ..................................................................................................................

\medskip
\noindent
{\em b})
\citet{Ho:2015} presented the metallicity gradient properties of 49 spiral galaxies with absolute magnitudes $-16 > M_B > -22$
 in the N2O2 (KD02) abundance scale. We note the good overlap in absolute magnitude with our LSBG sample.
 In the previous section we showed that when measuring gradient slopes, in alternative to the KD02 calibration, we can use the B07 calibration, with the results
summarized in Table~\ref{tab-gradients}. 
\citet{Ho:2015} found evidence for a common gradient among their galaxies, when the slope is expressed in units of the isophotal radius $r_{25}$.  
For the fainter half of their sample ($M_B > -20.1$), to which we  compare our galaxies, they found a mean slope $\alpha_{r_{25}} = -0.34$ dex\,$r_{25}^{-1}$, with a standard deviation of 0.17
(for the full sample the slope is similar:  $-0.39$ dex\,$r_{25}^{-1}$, with a standard deviation of 0.18). This compares to 
an average value  $\alpha_{r_{25}} = -0.133$ dex\,$r_{25}^{-1}$ for our LSBG sample, with a standard deviation of 0.072. The bottom panel of Fig.~\ref{histogram} shows the distribution of the gradient slopes 
for the \citet{Ho:2015} sample (black, dot-dashed line) and the LSBGs (red, continuous line).

The $t$--test indicates that in this case the difference between the average slopes  of HSBGs and LSBGs is  statistically  significant ($p = 0.0004$). A similar comparison with the HSBG sample presented by \citet{Pilyugin:2014}, excluding galaxies in pairs or mergers, as discussed by \citealt{Ho:2015}, yields the same conclusion.

\begin{figure}
\center
\includegraphics[width=0.47\textwidth]{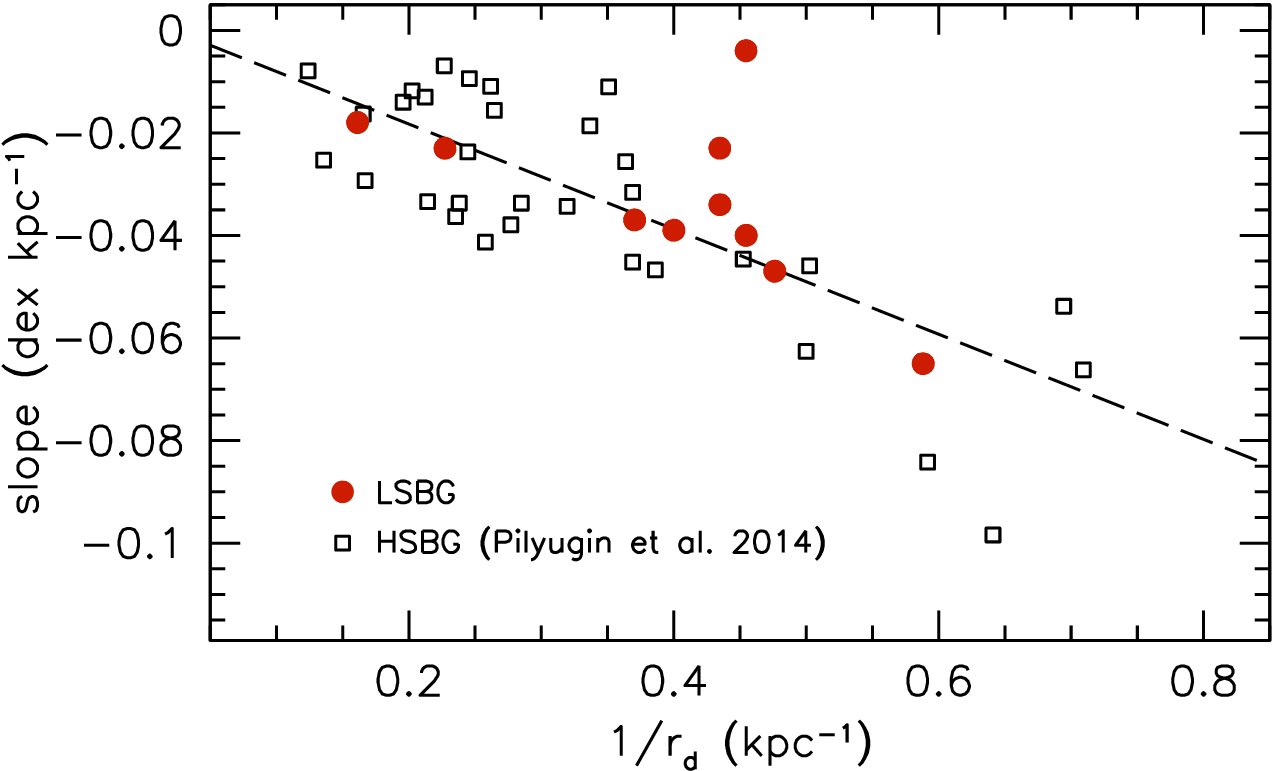}
\caption{Slope of the abundance gradients, in dex\,kpc$^{-1}$, as a function of the $B$-band exponential disc scale length $r_d$ for our LSBGs (red full dots) and a sample of HSBGs (open squares; slopes from 
\citealt{Pilyugin:2014} and disc lengths from \citealt{Pilyugin:2014a}). 
\label{rd}}
\end{figure}
% ..................................................................................................................

\subsection{Interpretation of the observed trends: clues on the common abundance gradients of spiral galaxies}
We show in Fig.~\ref{rd} the slope of the abundance gradients, now expressed in dex\,kpc$^{-1}$, as a function of the reciprocal of the  $B$-band exponential disc scale length $1/r_d$ for our LSBGs (red full dots) and a sample of HSBGs for which the relevant information is readily available
(open squares; slopes from 
\citealt{Pilyugin:2014} and disc lengths from \citealt{Pilyugin:2014a}). The plot illustrates how   LSBGs  follow the same trend observed for HSBGs, namely that the gradient slope in dex\,kpc$^{-1}$ anticorrelates with galaxy size: large discs display shallower gradients compared to small discs. As suggested by \citet{Prantzos:2000}, who discussed a similar plot in their modeling of spiral galaxies as a function of the rotational velocity and spin parameter of their haloes, this trend describes a fundamental property of galaxy discs, since 1/$r_d$ measures the radial decrease of the surface brightness with radius in mag\,arcsec$^{-2}$\,kpc$^{-1}$ [this follows from the equation for the exponential decrease of the surface brightness $\Sigma(r)=\Sigma(0) \exp(-r/r_d)$ and the corresponding 
equation when using magnitudes $\mu(r) = \mu_0 + 1.086\, r/r_d$]. Our plot in Fig.~\ref{rd} suggests that the underlying cause for the rate of radial decrease of the metallicity per kpc  in discs is  the same for all galaxies, regardless of their surface brightness. We argue that this is simply the rate of variation of the mass surface density with radius (also described by 1/$r_d$), through the  local mass surface density--metallicity relation $\Sigma_M-Z$ (\citealt{Edmunds:1984,Rosales-Ortega:2012, Moran:2012}), or equivalently the local stellar surface brightness--metallicity relation $\Sigma_L-Z$ (\citealt{Ryder:1995, Pilyugin:2014a}). 
Smaller galaxies have a steeper slope (in dex\,kpc$^{-1}$) in the radial decrease of both the surface mass density (or surface luminosity) and metallicity.
A linear least square fit to the data in Fig.~\ref{rd} yields the following:

\begin{equation}
\frac{d \log(O/H)}{dr} = -0.102\, (\pm 0.014) \frac{1}{r_d} + 0.002\, (\pm 0.005) \label{Eq-slope}
\end{equation}

\noindent
with distances expressed in kpc. Recalling that we can write $1/r_d = \Delta\mu/(1.086 \Delta r)$ using the variation of the surface brightness between two points separated 
by a linear distance $\Delta r$, we can derive from Eq.~(\ref{Eq-slope}) a simple relation between the change of $O/H$ and the corresponding variation in surface luminosity $\log\Sigma$: 

\begin{equation}
\Delta \log(O/H) = 0.235\,(\pm 0.032)\, \Delta \log\Sigma_{L_B} \label{Eq-trend}
\end{equation}

From the same HSBG galaxies included in Fig.~\ref{rd}
\citet{Pilyugin:2014a}  determined that  $\Delta\log(O/H) = 0.286\,(\pm 0.090)\, \Delta\log \Sigma_{L_B}$ for the centers of spiral discs, which agrees with Eq.~(\ref{Eq-trend}) within
the errors. Relations with comparable coefficients, valid across the discs of spirals, were found by \citet{Ryder:1995}, but for different photometric passbands (\eg\ $0.237 \pm 0.085$ in the $I$-band).

\medskip
If the underlying mechanism is the same why, then, is the mean slope of the gradient of LSBGs shallower than for HSBGs, when expressed in terms of the isophotal radius $r_{25}$, but similar
when referred to the exponential disc scale length, as shown above? Our interpretation, outlined below, is that the former result is due to a sort of selection effect, dependent on the disc central surface brightness. Our argument  also explains the existence of a common abundance gradient slope for HSBGs, as confirmed by \citet{Sanchez:2014} and others.

Let us consider abundance  gradients expressed in terms of the  isophotal radius, $r_{25}$. For a typical HSBG, we adopt the central surface brightness as given by the canonical \citet{Freeman:1970} value of $\mu_B(0) = 21.65$ mag\,arcsec$^{-2}$. Across the isophotal radius, the surface brightness varies by $\Delta\mu_{_{\!\mbox{\tiny\rm HSBG}}} = 25-21.65 = 3.35$ mag\,arcsec$^{-2}$. For a typical LSBG, we adopt instead $\mu_B(0) = 23.2$ 
mag\,arcsec$^{-2}$ (the median value in Table~1, but this choice is somewhat arbitrary), and the  variation over the isophotal radius is reduced to $\Delta\mu_{_{\!\mbox{\tiny\rm LSBG}}} = 1.8$ mag\,arcsec$^{-2}$. Converting surface brightness  to surface luminosity $\Sigma_L$ (\eg~measured in $L_\odot\,pc^{-2}$)
the corresponding radial variations are ($\Delta\log \Sigma_L)_{_{\!\mbox{\tiny\rm HSBG}}} = 1.34$ and ($\Delta\log \Sigma_L)_{_{\!\mbox{\tiny\rm LSBG}}} = 0.72$. We assume now that the  linear form of the local relation between log(O/H) and $\log \Sigma_L$ applies to all galaxies. We avoid the use of the local $\Sigma_M - Z$ relation, which requires an assumption about the mass-to-light ratio, and although these surface density--metallicity relations are better fit with second order polynomials (\citealt{Rosales-Ortega:2012})
or using additional parameters (\eg\ disc scale length or morphology, \citealt{Pilyugin:2014a}), the intent of our estimate is simply to reproduce the essential trends.

Using Eq.~(\ref{Eq-trend}) we can now estimate the variation of the O/H ratio across the isophotal radius $r_{25}$ to be $\Delta\log(O/H)_{_{\!\mbox{\tiny\rm HSBG}}} = 0.31$ dex  and $\Delta\log(O/H)_{_{\!\mbox{\tiny\rm LSBG}}} = 0.17$ dex. These estimates compare well with the actual average measurements discussed above ($0.34$ and $0.13$ dex, respectively), and the small differences  can easily be attributed both to the approximate nature of our calculations 
(including the choice of the central surface brightness for LSBGs) and the relatively large scatter in the observational data. We can conclude that LSBGs have a shallower mean abundance gradient in comparison to HSBGs, when expressed in terms of the isophotal radius, simply as a result of the reduced range in surface brightness  measured between their centers and their isophotal radii.
This effect is naturally not observed when the gradients are expressed in terms of the  exponential disc scale length because over such a length the variations in surface brightness, surface luminosity and surface mass density are independent of the central surface brightness (e.g.~the surface brightness decreases by a constant  1.086 magnitudes arcsec$^{-2}$ over this length).

\medskip
A far-reaching corollary to these simple considerations is that the common abundance gradient slopes observed in HSBGs, expressed either in terms of disc scale lengths or  isophotal radii (as found 
by \citealt{Sanchez:2014}, \citealt{Ho:2015} and others), and their relatively small dispersions, derive naturally from the surface mass density (or luminosity)--metallicity relation (see also the discussion in \citealt{Sanchez:2014}) and, in the case of the gradients per isophotal radii, the narrow range in central surface brightness spanned by the HSBGs typically selected for chemical abundance analysis. With a dispersion  of 0.3 mag\,arcsec$^{-2}$ in  $\mu_B(0)$ (\citealt{Freeman:1970}) and a dispersion of 0.11 dex in O/H coming from the empirical  $\Sigma_L - Z$ relation (\citealt{Pilyugin:2014a}) we roughly estimate a dispersion in the gradient slope of $\sim 0.12$ dex\,r$_{25}^{-1}$, which compares well with $\sigma = 0.12$ dex\,r$_{25}^{-1}$ and $\sigma = 0.18$ dex\,r$_{25}^{-1}$
reported by \citet{Sanchez:2014} and \citet{Ho:2015}, respectively. 
These figures can be improved with appropriate modeling of the uncertainties involved, but this lies beyond the scope of our study.
Considering scale lengths instead, the mean value found in the $g$-band by \citet{Fathi:2010} for a large sample of galaxies from the Sloan Digital Sky Survey is $3.85 \pm 2.10$~kpc. Using Eq.~(\ref{Eq-slope})
we derive a corresponding gradient of $\alpha = -0.15$ dex\,$r_e^{-1}$, with a standard deviation of 0.08 dex\,$r_e^{-1}$. There is a systematic offset in comparison with the average slope measured by
\citet{Sanchez:2014}, but the standard deviation is the same.

Lastly, we  conclude that the common abundance gradient measured in dex $r_d^{-1}$ expresses a more fundamental property of spiral discs than the common gradient  in dex $r_{25}^{-1}$,
being linked to the underlying physics (the existence of a local mass surface density--metallicity relation and its relative variations with radius) in a simpler way, without dependence on  structural properties such as the actual value of the central mass surface density.

%==============================================================================================================
\section{Summary}

Our observations show that LSBGs  have measurable radial chemical abundance gradients, qualitatively similar to those present in the  more extensively studied high surface brightness galaxies.
This result contrasts with the previously held notion that LSBGs do not possess such gradients. 
In addition, we found that the mean LSBG gradient, when expressed in terms of the isophotal radius, is significantly shallower than for the average HSBG. Since 
both LSBGs and HSBGs follow the same relation between the radial gradients of both  metallicity (in dex\,kpc$^{-1}$) and  surface brightness (in magnitudes\, arcsec$^{-2}$\,kpc$^{-1}$), we can explain this result simply adopting a common surface luminosity--metallicity relation (reflection of the more fundamental relation involving the mass surface density)
and considering the range in surface brightness  across the optical discs.
The common gradient in dex\,r$_{25}^{-1}$
observed in HSBGs derives from  the similarity of their surface brightness (or equivalently mass surface density) gradients and central surface brightness. 
The latter conclusion does not contradict alternative explanations for the existence of this common abundance gradient. For example, \citet{Ho:2015} based their interpretation on the
similarity of the radial gas and stellar  surface density  profiles among spiral discs. 
Our reformulation, based on the structural similarity  and the surface brightness properties, yields a concordant, equivalent interpretation.
On the other hand, when normalizing the gradients to the exponential disc scale lengths we directly expose the underlying physical mechanism, i.e.~the mass surface density--metallicity
relation, and the homology with which it varies with galactocentric distance among galaxies.

% ..................................................................................................................

\bigskip
\bigskip
\noindent

\section*{Acknowledgments}
Based on observations collected at the European Southern Observatory, Chile, under programs 386.B-0144 and 089.B-0351, and at the Gemini Observatory, which is operated by the 
Association of Universities for Research in Astronomy, Inc., under a cooperative agreement 
with the NSF on behalf of the Gemini partnership: the National Science Foundation (United 
States), the Science and Technology Facilities Council (United Kingdom), the 
National Research Council (Canada), CONICYT (Chile), the Australian Research Council (Australia), 
Minist\'{e}rio da Ci\^{e}ncia e Tecnologia (Brazil) 
and Ministerio de Ciencia, Tecnolog\'{i}a e Innovaci\'{o}n Productiva (Argentina). 

%\clearpage
\bibliographystyle{mnras}
\bibliography{References}

\bsp	% typesetting comment
\label{lastpage}
\end{document}